\documentstyle[11pt]{article}

\catcode`\@=11
\def\theequation{\arabic{equation}}

\def\elab#1#2{\begin{equation}\label{e:#1}#2\end{equation}}
\def\eref#1{Eq.~\ref{e:#1}}
\def\eqsref#1{Eqs.~\ref{e:#1}}

\def\ealab#1#2{\begin{eqnarray}\label{e:#1}#2\end{eqnarray}}
\def\beq{\begin{equation}}
\def\eeq{\end{equation}}

\def\fig#1#2#3{\begin{figure}[htbp]
  \vspace{#3}
  \caption[dummy]{#2}
  \label{f:#1} \end{figure}}

\def\fref#1{Fig.~\ref{f:#1}}

\catcode`\@=11
\def\rref#1{Ref.~{\csname b@r:#1\endcsname}}
\catcode`@=12

\def\ie{\hbox{\it i.e.}{}}      \def\etc{\hbox{\it etc.}{}}
\def\eg{\hbox{\it e.g.}{}}      
\def\etal{\hbox{\it et al.}}

\def\deriv#1#2{\frac{d#1}{d#2}}

\def\pfrac#1#2{ \left( \frac{#1}{#2} \right)}

\def\simlt{\ \raisebox{-.25ex}{$\stackrel{<}{\scriptstyle \sim}$}\ }

\def\VEV#1{\left\langle #1\right\rangle}

\def\abs#1{\left| #1\right|}
\def\Hefour{{}^4{\rm He}}
\def\Hethree{{}^3{\rm He}}

\def\Liseven{{}^7{\rm Li}}

\def\anti#1{\overline#1}

\def\nue{\nu_e}

\def\nuebar{\anti{\nu}_e}

\def\Gfermi{G_f}
\def\thetac{\theta_c}

\def\en#1#2{{#1} \times 10^{#2}}
\def\M#1{^{-#1}}
\def\UN#1{\ {\rm #1}}

\def\GeV{\UN{GeV}}
\def\MeV{\UN{MeV}}

\def\keV{\UN{keV}}

\def\sec{\UN{sec}}

\catcode`@=11
\def\@cite#1#2{$^{\,#1\if@tempswa , #2\fi}$}
\def\@citex[#1]#2{\if@filesw\immediate\write\@auxout{\string\citation{#2}}\fi
  \def\@citea{}\@cite{\@for\@citeb:=#2\do
    {\@citea\def\@citea{,}\@ifundefined
       {b@\@citeb}{{\bf	?}\@warning
       {Citation `\@citeb' on page \thepage \space undefined}}%
{\csname b@\@citeb\endcsname}}}{#1}}
\catcode`@=12

\def\CI#1{\cite{r:#1}}
\def\BI#1{\bibitem{r:#1}}

\def\PRD#1#2#3#4{#1, {\it Phys. Rev.} {\bf D#2}, #3 (19#4)}

\def\PLB#1#2#3#4{#1, {\it Phys. Lett.} {\bf #2B}, #3 (19#4)}
\def\ApJ#1#2#3#4{#1, {\it Astrophys. J.} {\bf#2}, #3 (19#4)}

\def\NPA#1#2#3#4{#1, {\it Nucl. Phys.} {\bf A#2}, #3 (19#4)}
\def\NPB#1#2#3#4{#1, {\it Nucl. Phys.} {\bf B#2}, #3 (19#4)}

\def\NIM#1#2#3#4{#1, {\it Nucl. Instrum. Methods}, {\bf#2}, #3 (19#4)}

\def\PRep#1#2#3#4{#1, {\it Phys. Rep.} {\bf#2}, #3 (19#4)}

\def\SJNP#1#2#3#4{#1, {\it Sov. Jour. Nucl. Phys.} {\bf#2}, #3 (19#4)}

\def\baselinestretch{1.5}

\def\yhe{Y_4}
\def\mn{m_n}
\def\mp{m_p}
\def\mpi{m_\pi}
\def\mnuc{m_N}
\def\xn{x_n}
\def\xp{x_p}
\def\pn{p \leftrightarrow n}
\def\pton{p \rightarrow n}
\def\ntop{n \rightarrow p}
\def\Td{T_d}
\def\Tf{T_F}
\def\Tgamma{T_\gamma}
\def\Tnu{T_\nu}
\def\me{m_e}
\def\taun{\tau_n}
\def\taunrec{\tau_{n,rec}}
\def\nun{\nue n \rightarrow e^- p}
\def\ep{e^- p \rightarrow \nue n}
\def\nubarp{\nuebar p \rightarrow e^+ n}
\def\eplusn{e^+ n \rightarrow \nuebar p}
\def\ndec{n \rightarrow \nuebar e^- p}
\def\invdec{e^- \nuebar p \rightarrow n}
\def\vrel{v_{rel}}
\def\vnuc{v_{nuc}}
\def\fudge{f_\alpha}
\def\ctheta{\cos\theta}
\def\calpha{\cos\alpha}
\def\fwm{f_2}
\def\fps{f_{ps}}
\def\fb{{\rm f_b}}
\def\fbiso{{\rm f_{b,1}}}
\def\fbtheta{{\rm f_{b,\theta}}}
\def\fbalpha{{\rm f_{b,\alpha}}}
\def\fbalphatheta{{\rm f_{b,\alpha\theta}}}
\def\ff{\vrel'}
\def\sigmawm{\sigma_{wm}}

\def\gammawm{\gamma_{wm}}
\def\gammarec{\gamma_{rec}}
\def\gammafb{\gamma_{\fb}}
\def\gammatha{\gamma_{th,0}}
\def\gammathb{\gamma_{th,\theta}}
\def\gammath{\gamma_{th}}
\def\gammadec{\gamma_{\ndec}}
\def\gammaalpha{\gamma_{\alpha}}
\def\gammaalphatheta{\gamma_{\alpha,\theta}}
\def\gammaalphatot{\gamma_{\alpha,tot}}
\def\betacm{\beta_{cm}}
\def\gammacm{\gamma_{cm}}
\def\gampn{\Gamma_{\pton}}
\def\gamnp{\Gamma_{\ntop}}
\def\resid{\delta_2}
\def\deltasc{\delta_{sc}}
\def\deltan{\delta_{n}}
\def\delyhe{\Delta \yhe}
\def\delxn{\Delta \xn}
\def\Ecm{E_{3,cm}}
\def\kcm{k_{3,cm}}

\begin{document}

\begin{titlepage}
\begin{flushright}
BA-93-16 \\
hep-ph/9305311 \\
\today
\end{flushright}
%\par
\vspace{.3in}
\begin{center}
{\Large{\bf Nucleon Mass Corrections to the $\pn$ Rates
During Big Bang Nucleosynthesis}}\\
\vskip 0.3in
{\bf David Seckel}\\[.2in]
{\it Bartol Research Institute \\
University of Delaware, Newark, DE 19716\\}

\end{center}
\vskip 0.4in

\begin{abstract}
The thermal rates for converting neutrons to protons, and vice versa, are
calculated, including corrections of order $1 \MeV$ divided by a nucleon
mass. The results imply that the primodial helium abundance predicted
for big bang nucleosynthesis has been systematically underestimated
by about $\delyhe = 0.0012$, \ie, $\delyhe / \yhe \approx .005$.
\end{abstract}

%\begin{flushleft}
%Pacs numbers:  14.80.Pb, 97.60.Bw, 25.90.+k
%\end{flushleft}
\end{titlepage}

\def\baselinestretch{1.0}

\section{Introduction}

The purpose of this paper is to evaluate nucleon mass corrections to the
rate of weak transitions that interconvert neutrons and protons
during the early stages of big bang
nucleosynthesis\cite{r:KolbT,r:WalkerSSOK}.
In the usual calculation of these rates the nucleon mass is ignored; \ie\
one includes all energies and momenta
in the\MeV\ range, specifically, ratios of the electron mass, $\me$, the
temperature, $T$, and the neutron-proton mass difference\CI{pdb},
$Q = \mn - \mp = 1.2933 \MeV$; but factors such as $Q/\mnuc$, \ie, a low energy
scale divided by a nucleon mass, are ignored. These factors are
individually of order a tenth of a percent, but it will be shown that
together they cause roughly a 0.5\% increase in the helium abundance predicted
by big bang nucleosynthesis calculations. Such a systematic correction is
significant in that it is comparable to the
largest uncertainty in the standard hot big bang
calculation - that due to uncertainty in the neutron half life. Further,
the increase in the predicted helium abundance translates into a tighter
constraint on the density of baryons as well as a strengthening of
particle constraints based on big bang nucleosynthesis - such as
the limit on the number of neutrino species that may be in thermal
equilibrium in the early Universe\CI{SteigmanSG}.

As an example of the sort of effect that is usually ignored,
consider the neutron to proton abundance ratio in thermal equilibrium.
Including the first correction in an expansion in inverse powers of $\mnuc$
this ratio is
\elab{npratio}{
 \frac{\xn}{\xp} \approx e^{-Q/T} \pfrac{\mn}{\mp}^{3/2}
     \approx e^{-Q/T} (1 + 1.5 \frac{Q}{\mnuc}) = 1.00207 e^{-Q/T} .
}
Usually one includes just the `Boltzman factor' and ignores the
correction, which is small, $0.2\%$. Thus, even if
freeze out of the weak reactions occurred at the same time, one might
expect the neutron abundance to be slightly higher if nucleon mass
corrections were included.

Of course, it is essential that the neutron fraction drops out of
thermal equilibrium as the weak reactions become slow compared to the
expansion rate of the Universe, and so one does not calculate the
neutron abundance by equilibrium arguments in a numerical calculation.
Instead one evaluates the rates for $\pn$ conversions and tracks carefully
the maintenance of approximate equilibrium at high temperatures and the
failure to maintain equilibrium at low temperatures. It is, therefore,
necessary to evaluate the change in the rates - not just the equilibrium
neutron fraction. There are many $1/\mnuc$ corrections to the $\pn$ rates
and it is the purpose of this paper to enumerate and evaluate them
in a systematic fashion.

The spirit of this paper is similar to those which
evaluated the electromagnetic radiative, thermal, and coulomb
corrections to the $\pn$ processes\cite{r:DicusKGSTT,r:CambierPS}.
In both cases, the corrections are a few percent at most.
To achieve a satisfactory level of
accuracy, one part in a thousand, it is necessary to evaluate only the
first correction, but not effects of order $1/\mnuc^2$ or, in the
electromagnetic case, of order $\alpha^2$. Nor is it necessary to
consider terms of order $\alpha /\mnuc$.

Another similarity in the two problems concerns the normalization
of the corrections. The nucleosynthesis numerical codes typically
normalize the weak rates to the experimental value of the neutron mean
lifetime, $\taun = 889.1 \pm 2.1 \sec$. Thus, when evaluating a purported
correction to the rates one must also evaluate the same sort of
corrections for neutron decay, and adjust the corrections appropriate for
BBN accordingly. So, for example, the
largest term in the order $\alpha$ radiative correction to the weak rates is
a constant which also shows up in neutron decay. Similarly, a good part of
the coulomb correction to the weak rates also cancels. Thus, the early
numerical code of Wagoner\cite{r:Wagoner,r:KawanoS}
contained a simple coulomb correction and no
radiative correction, but although individual reactions have corrections
of $\sim 5\%$, the net effect of a more detailed treatment results in less than
a 1\% correction to Wagoner's results. In contrast, the $1/\mnuc$
corrections are of order 1\% to the reaction rates, but the comparable
correction to neutron decay is smaller due to kinematic thresholds.
As a result, nearly the whole of the effects discussed here survive
to affect the helium abundance.

With these thoughts in mind the rest of the paper is ordered as follows.
In section 2, the main results are presented - the corrections to the
$\pn$ rates to first order in $1/\mnuc$. In section 3, similar effects are
considered for neutron decay. Section 4 combines the results from the
previous two sections to arrive at an expected change in the helium
abundance. Section 5 contains a discussion of the significance of the
results.

First, however, it may be useful to the reader to clarify some of the
notation used later. Except where the neutron or proton mass is explicitly
indicated by $\mn$ or $\mp$, the nucleon mass is given as $\mnuc$.
In the formulae for cross-sections, rates, \etc, $\mnuc$ refers to
the initial nucleon mass, but to the extent that the formulae are only
accurate to $1/\mnuc$ it makes no difference which nucleon mass is
actually used. Also, $E_1$ and $k_1$ are the energy and momentum
of the initial lepton in the rest frame of the fluid. Unless specifically
noted, the energy $E_3$ denotes the quantity $E_1 + dQ$, where $dQ = \pm Q$.
This is only equal to the outgoing lepton energy in the infinite mass
limit, $\mnuc \rightarrow \infty$. $k_3$ is the corresponding momentum,
$k_3 = (E_3^2 + m_3^2)^{1/2}$. During nucleosynthesis the temperature
describing the neutrino distribution, $\Tnu$, is not equal to the
temperature of the rest of the plasma (including the nucleons),
denoted by $\Tgamma$. When the temperature $T_3$ is used, it refers to the
temperature which describes the outgoing lepton.

\section{Corrections to Scattering Processes}

There are six processes that contribute to $\pn$ conversion in the early
Universe; neutron decay $\ndec$, inverse neutron decay $\invdec$,
and four scattering processes, $\nun,\ \ep,\ \nubarp$, and $\eplusn$. The most
critical time is when these reactions are `freezing out', \ie, when they
are just failing to maintain thermal equilibrium. This occurs at a
temperature $\Tf \approx 0.8 \MeV$. At that time the scattering processes
dominate over neutron decay and inverse decay by a factor of about 1000.
The reactions which convert neutrons to protons are some 6 times greater
than the inverse reactions, due to the nuclear mass difference affecting
phase space. The reactions involving antileptons are nearly equal in
importance to those involving leptons. To achieve an accuracy of 0.1\% it
therefore seems sufficient to consider just corrections to
the scattering rates;
however, because of the role played by neutron decay in normalizing the
weak rates those corrections must also be examined. Accordingly,
this section presents corrections to scattering and the next examines
neutron decay.

The rate for two body scattering reactions in a medium may be
written in the form
\elab{2to2a}{
\Gamma(12 \rightarrow 34) =
\left(\prod_i \frac{\int d^3 k_i}{(2 \pi)^3 2 E_i} \right)
  (2 \pi)^4 \delta^4(\hbox{$\sum_i$} p_i) \abs{\cal{M}}^2
    n_1 n_2 (1 - n_3) (1 - n_4),
}
where $p_i$ is the four momentum, $k_i$ is the three momentum, $E_i$
is the energy of each particle, and for the problem at hand all
particles obey Fermi statistics. The occupation numbers $n_i$ take
thermal equilibrium values only for those species actually in equilibrium.
In this notation the squared matrix element, $\abs{\cal{M}}^2 $,
has been summed over all spin
degrees of freedom and it is assumed that the $n_i$ do not depend on
spin. The presence or absence of right-handed neutrinos is irrelevant for
the evaluation of the scattering rates. For the reactions of interest,
let particles 2 and 4 be the in and out nucleons, respectively,
and let particles 1 and 3 be the leptons or anti-leptons.

\subsection{The infinite mass limit}
Before going into the details of nucleon mass corrections it is appropriate
to evaluate the reaction rates in the limit of infinite nucleon mass.
In this limit the energies of the in and out leptons are related
by $E_1 + dQ = E_3$, where $dQ = \pm Q$ depending on
whether a neutron or proton is the initial nucleon. Further,
neither lepton occupation number will depend upon the scattering
angles. The rate can then be rewritten in the familiar form
\elab{2to2b}{
\Gamma(12 \rightarrow 34) =
  \int \frac{d^3 k_1}{(2\pi)^3} \frac{d^3 k_2}{(2 \pi)^3}
  \sigma \vrel n_1 n_2 (1 - n_3) (1 - n_4),
}
where $\sigma$ is the cross-section for the reaction summed over both
initial and final spins and $\vrel$ is the
relative velocity of the two initial particles, which for infinite mass
nucleons may be taken to be just the initial lepton velocity, $v_1$.
It is useful to concentrate on the rate per initial state in the
absence of blocking, $\gamma \equiv \sigma \vrel$. For infinite mass nucleons,
this quantity becomes
\elab{sigma0}{
\sigma\vrel \rightarrow \fudge (\sigma_0 v_1) = \fudge \gamma_0
  \frac{2 \Gfermi^2 \cos^2\thetac (1 + 3 c_a^2)}{\pi} E_3 k_3 =
   \fudge A E_3 k_3,
}
where $\sigma_0$ is the cross-section for a lepton of energy $E_1$
incident on an infinitely heavy nucleon,
$\Gfermi = \en{1.1164}{-5}\GeV\M2$ is Fermi's constant,
$\thetac$ is the Cabibbo angle, with $\cos\thetac = .975$,
and $c_a \equiv g_A/g_V = 1.257$ is the ratio of the axial vector to vector
coupling of the nucleon for charged currents. The last relation in
\eref{sigma0} serves as a
definition of the constant $A$, a factor which will be common to all the
weak reactions. The coulomb and radiative corrections to the rates
are embodied in an electromagnetic correction factor $f_\alpha$; however,
as explained in the introduction,
all corrections to the weak rates are small and may be treated independently.
It is therefore acceptable to ignore $\fudge$ except when worrying about
the overall normalization of the rates, and so the factor $\fudge$ will be
dropped.

For heavy nucleons, low baryon density and low lepton asymmetry, it is
appropriate to approximate $n_2$ by a Boltzman distribution and
ignore $n_4$ entirely. Integrating over nucleon momentum and lepton
direction, converting the lepton momentum integral to
one over energy, and using thermal distributions for the leptons
one gets the rate,
\ealab{2to2c}{
\Gamma(12 \rightarrow 34) & = & \int d E_1 \deriv{\Gamma}{E_1} \nonumber \\
& = & \frac{N_2}{4 \pi^2}
  \int d E_1 \frac{E_1 k_1 \gamma_0}{(1 + e^{E_1/T_1})(1 + e^{-E_3/T_3})}
   \nonumber \\
& = & \frac{A N_2}{4 \pi^2}
  \int d E_1 \frac{E_1 k_1 E_3 k_3}{(1 + e^{E_1/T_1})(1 + e^{-E_3/T_3})},
}
where $N_2$ is the spatial density of initial nucleons,
and $T_i$ is the temperature describing lepton $i$. \eref{2to2c} defines
the differential interaction rate $\deriv{\Gamma}{E_1}$, which is plotted
in \fref{tot0} for each of the four scattering processes. The plots were
generated using temperatures $\Tgamma = 0.8 \MeV$,
and $\Tnu = 0.07926\MeV$. This is near
the conventional ``freezeout temperature'', \ie\ that temperature where
the equilibrium abundances are equal to the final values, as if the
interactions were very rapid and then turned off abruptly. The freezeout
point is high enough that electron annihilation has caused the photon
temperature to increase by only a small amount over the neutrino
temperature. Note that the $\ntop$ rates in \fref{tot0} are some 6 times
greater than the $\pton$ rates, as required to maintain equilibrium at
this temperature.

To give a better feel for the important points in determining
the neutron fraction, \fref{inttot0} shows the integrated scattering rates,
$\Gamma$, for the four scattering processes as a function of the photon
temperature, $\Tgamma$. The expansion rate, $H(\Tgamma)$, is also shown;
along with the free neutron decay rate. The $\pton$ reactions freezeout
first, and become increasingly unimportant at lower temperatures. The
$\ntop$ scattering rates freeze out later. They are more important than
free neutron decay down to a temperature $\sim 0.2\MeV$, but what really
counts is the comparison to $H$. After $\Tgamma \simeq 0.5\MeV$
the most significant comparison to $H$ is free neutron decay
just at the time the ``deuterium bottleneck'' breaks, at which time
$\Tgamma \approx 0.07 \MeV$. Corrections to the
scattering rates for $\Tgamma \simlt 0.5 \MeV$ are not very important.

Apart from electromagnetic corrections, \eref{2to2c} is the reaction rate
used in nucleosynthesis calculations. There are several points where
infinite mass nucleons were used. Merely writing the reaction in the form of
a cross-section required that the final state occupation numbers did not
depend on the scattering angles, and this depends on the assumption that no
recoil energy goes to the nucleon. The vector and axial vector
cross-section, \eref{sigma0}, has corrections of order $1/\mnuc$. In addition,
the vector and axial vector Lagrangian must be corrected for nucleon
structure effects, such as momentum dependent form factors, or new terms
in the effective low energy effective Lagrangian, such as `weak
magnetism'. The relative velocity, $\vrel$, must be corrected for the
nucleon velocity. One must average over the Boltzman distribution, $n_2$, to
produce a `thermal averaged' cross-section $\times$ relative velocity
$\times$ blocking effects due to the Fermi statistics. Since the nucleon
velocity $\vnuc$ is of order $(\Tgamma/\mnuc)^{1/2}$ one must expand in the
nuclear
velocity to second order to get corrections to first order in the nucleon
mass. One implication of this is that there may be correlations between
corrections that are first order in $\vec{v}_{nuc}$. Although first
order terms vanish when angle averaged their correlations may not, and
can therefore contribute at first order in $1/\mnuc$.

As presented here, these calculations are done by evaluating corrections to
the rates, $\gamma = \sigma \vrel$, and the blocking factors $\fb = 1-n_3$,
as a function of the initial lepton energy. After taking appropriate
angular combinations, the corrections are
integrated over lepton energy to produce corrections to the conversion
rates per nucleon, which may be used in the rate equations to solve for
the neutron abundance as a function of time.

The most obvious corrections to consider are $1/\mnuc$ corrections
to the cross-section, which are combined with the zeroth order, or
infinite nucleon mass, values for $\vrel$ and $\fb$.
The $1/\mnuc$ corrections to $\sigma$ will be calculated in section 2.2.
When the nucleon velocity is taken into account there will be corrections
to $\sigma$ due to the altered lepton energy, as well as corrections
intrinsic to $\vrel$. These
are evaluated in section 2.3 along with the corrections to the blocking
factors. These corrections and their correlations will be expressed as
effective corrections to the rate $\gamma_0$, which can be multiplied by
the zeroth order $\vrel$ and $\fb$.

As a preliminary to this, consider the differential
cross-section to order $1/\mnuc$,
\elab{sigma}{
4 \pi \deriv{\sigma}{\Omega} = \sigma_0 + \sigma_1
  + \sigma_\alpha \calpha.
}
The zeroth order total cross-section, $\sigma_0$, is
given by \eref{sigma0}. The $1/\mnuc$ correction to the total
cross-section is $\sigma_1$. The relevant angular dependence of the
differential cross-section is given by $\sigma_\alpha$, which
is to be multiplied by $\calpha$ with $\alpha$ being the center of
mass scattering angle. Terms higher order in $\calpha$ are suppressed
by two powers of $\mnuc$ (or powers of $\Gfermi$) and may be dropped.
If there were no corrections to the blocking factors
for the final state leptons $\sigma_\alpha$ would integrate to
zero when averaged over scattering angle; however, there are corrections
to the blocking factors. Since these are
suppressed by factors of $\mnuc$ one need only keep the zeroth order term
of $\sigma_\alpha$,
\elab{sigmaalpha}{
\sigma_\alpha = \sigma_0 \frac{1 - c_a^2}{1 + 3 c_a^2}
  \frac{k_1 k_3}{E_1 E_3}.
}
Discussion of the corrections to the rates due to $\sigma_\alpha$
is postponed till later, after evaluating the
corrections to the lepton blocking factors in section 2.3.

\subsection{Corrections to the cross-section}

There are two important corrections in
$\sigma_1$, one that arises from including the weak magnetism term in the
interaction, and one that arises from modifications to the final state
phase space due to the recoil of the nucleon. They may be treated
independently to first order in $\mnuc$.

The effective low energy weak Lagrangian is,
\elab{lweak}{
{\cal L}_w = \frac{\Gfermi}{\sqrt{2}} J_{lep}^\mu J_{had,\mu},
}
where the leptonic current has the usual $V- A$ structure and
the hadronic weak current is given by\CI{ComminsB}
\elab{hadcur}{
J_{had}^\mu = \cos\thetac
  \anti\psi_N \left( \gamma^\mu(1 - c_a \gamma_5) +
      i \frac{\fwm}{\mnuc} \sigma^{\mu\nu} q_\nu +
         \fps \gamma_5 q^\mu \right) \psi_N,
}
where $\fwm = 1.81$ is the anomalous weak charged current
magnetic moment of the nucleon,
$\fps$ is the pseudoscalar coupling to the nucleon, and $q$ is the
momentum transfer to the nucleon. At higher energies, one would treat the
couplings $g_A$, $g_V$, $\fwm$, and $\fps$ as form factors\CI{LlewellynSmith}
with corrections of order $q^2/M_i^2$, where the $M_i$ differ for the different
interactions and are experimentally determined to be in the range
$500 - 1000 \MeV$. Thus, the form factor corrections may reasonably be
assumed to be higher order than the $1/\mnuc$ corrections considered in
this paper.

The full squared matrix element for scattering with the current
in \eref{hadcur} is given in Appendix A, but here only the relevant terms
are kept. The pseudoscalar coupling is usually
approximated by the pion pole term. At low momentum transfer this leads
to a suppression of the amplitude by a factor of
$\sim g_{\pi NN} \me q^2 /(\mpi^2 f_\pi)$,
where $g_{\pi NN}$ is the $\pi$-nucleon coupling and $f_\pi$ is the pion
decay constant. Since this is small it is dropped
from further discussion. Weak magnetism is generated by the $\fwm$ term. There
is an explicit factor of $1/\mnuc$ in the coupling, so one may ignore the
square
of the weak magnetism term, but there may be interference
between weak magnetism and the vector and axial vector interactions.
The interference with the vector interaction vanishes at order $1/\mnuc$,
which leaves just a correction proportional to $c_a \fwm$,
\elab{sigmawm}{
\gammawm = \sigmawm v_1 = \gamma_0  \frac{c_a \fwm}{1 + 3 c_a^2}
  \frac{2 (E_3 k_1^2 + E_1 k_3^2)}{E_1 E_3 \mnuc},
}
where the zeroth order $\vrel$ is acceptable since there is already one
power of $1/\mnuc$ in the correction $\sigmawm$.

Next, consider the $1/\mnuc$ corrections to the usual $V$ plus $A$
interactions. These will be referred to collectively as
the `recoil' correction, since a
major component of the correction may be understood as a reduction in the
phase space for the outgoing lepton due to the energy carried off by the
nucleon. The correction is calculated in the frame of the target nucleon
by 1) expressing the differential cross-section in terms of
the invariants $s$, $t$, and the particle masses, 2) expressing $s$ and $t$
in terms of the incident lepton energy, 3) integrating over phase space,
and 4) extracting all terms to the required power of $1/\mnuc$. One must
keep the full expression for $s$, $s = m_2^2 + 2 E_1 m_2 + m_1^2$, since the
leading part of $s$ cancels in some parts of the calculation. The invariant
$t$ may be written as $t = t_0 + \delta t \calpha$, where to first order in
$1/\mnuc$
\ealab{trefs}{
t_0 & = & -2 E_1 E_3 + m_1^2 + m_3^2 +
  \frac{2 k_1^2 E_3 + E_1(k_1^2 + k_3^2)}{\mnuc} \nonumber \\
\delta t & = & 2 k_1 k_3 -
  \frac{k_1(2 E_1 k_3^2 + E_3(k_1^2 + k_3^2))}{k_3 \mnuc}.
}
Integrating over $dt$, keeping just the term proportional to $\mnuc$, and
applying the appropriate normalization yields
\ealab{sigmarec}{
\gammarec & = & \frac{1}{\fwm} \gammawm -
  \gamma_0 \frac{(2 E_1 k_3^2 + E_3 (k_1^2 + k_3^2))}{2 k_3^2 \mnuc} +
  \gamma_0 \frac{1}{1 + 3 c_a^2} \frac{(m_1^2 - m_3^2 - Q^2)}{2 E_3 \mnuc}
\nonumber \\
& & + \gamma_0 \frac{c_a^2}{1 + 3 c_a^2}
   \frac{(6 E_1^2 E_3 - 6 E_1 E_3^2 - 3 E_1 k_1^2
        - 4 E_3 k_1^2 - E_1 k_3^2)}{2 E_1 E_3 \mnuc}.
}
Note that the interference between the $A$ and $V$
currents has exactly the same structure as that between the axial
vector and weak magnetism interactions.

\subsection{Thermal averages of $\sigma \vrel$ and $\fb$}

For the remaining corrections one must perform averages over scattering angle
and/or thermal averages over the nucleon momentum. The strategy presented
here is to evaluate these corrections separately for the lepton blocking
factor $\fb$, and for the product $\sigma \vrel$. Each is developed as
a power series in the cosines of the scattering
angle $\alpha$ and of the incident angle of the initial lepton momentum
relative to the nucleon momentum, labeled by $\theta$. It is only
necessary to include terms up to $\cos^2 \theta$, since each factor of
$\ctheta$ comes accompanied by the nucleon velocity, which is of
order $\sqrt{\Tgamma/\mnuc}$. Further, terms first order in $\ctheta$ and
$\calpha$ integrate to zero and may be dropped, although only after
the two series are multiplied together to pick up the angular
correlations between the corrections to $\sigma \vrel$ and $\fb$.

The thermal averaged $\sigma \vrel$ for a lepton of energy $E_1$
is given by
\elab{thermave}{
\VEV{\sigma \vrel} = \frac{\int k_2^2 d k_2 d(\ctheta)}{4 \pi^2}
   \sigma \vrel n_2.
}
\eref{sigma0} can be used for $\sigma \vrel$ with just two
changes. First, one must use the lepton energy in the nucleon rest frame,
\elab{e1}{
E_1' = E_1 \gamma (1 - v_1 \vnuc \ctheta),
}
where here $\gamma = 1/(1-\vnuc^2)^{(1/2)}$ is the relativistic $\gamma$
factor for the initial nucleon. Second, $\sigma \vrel$ must be multiplied
by a factor
\elab{fluxfac}{
\ff = 1 - v_1 \vnuc \ctheta,
}
to account for the change in lepton flux seen in the nucleon rest frame.
The thermal average is then done by expanding in powers of $\vnuc$ and
replacing $\vnuc^2$ by its thermal average,
$\vnuc^2 \rightarrow 3 \Tgamma/\mnuc$.
This procedure is totally equivalent to the more standard practice of
using the Lorentz invariant cross-section with $E_1'$ and
using the Lorentz invariant flux factor
\elab{vrel}{
\vrel = \left( (\vec{v}_1 - \vec{v}_{nuc})^2 -
   (\vec{v}_1 \times \vec{v}_{nuc})^2 \right)^{1/2}.
}
However, a difficulty arises in the use of \eref{vrel}. When $\vrel$ is
expanded in terms of $\vnuc$ terms of order $\vnuc/v_1$ are generated,
but there is a region of phase space where the lepton velocity is small
compared to $\vnuc$, and this expansion is not valid. Using the rest frame
$\sigma \vrel$ and \eref{fluxfac} avoids this problem for the incident
lepton velocity.

The result of performing the thermal average is an effective correction to
$\sigma \vrel$ for incident lepton energy $E_1$,
\elab{sigmatha}{
\gammatha = \gamma_0 \frac{T}{\mnuc}
\left(
\frac{3 E_1^2 + 2 k_1^2}{2 E_1 E_3} +
\frac{3 k_1^2 E_1 + 3 E_1^2 E_3 + 2 k_1^2 E_3)}{2 E_1 k_3^2}
- \frac{k_1^2 E_3^2}{2 k_3^4}	\right).
}
The last term in \eref{sigmatha} presents a problem akin to that just
discussed concerning $\vrel$; namely, when the reaction energy is near
threshold the final state lepton velocity will be small if that lepton is
massive, \ie\ it is an electron or positron. This is not a problem for the
reaction $\nun$ since the positive $Q$ value always keeps the electron
energy well above threshold, but it is a problem for $\nubarp$. It is
also not a difficulty for reactions with final state neutrinos since then
$k_3 = E_3$.

The anomalous powers of $k_3$ are symptomatic of a deeper
problem with the thermal averaging. The averaging procedure adopted here
is only valid when the change in outgoing lepton momentum due to nuclear mass
effects is small compared to its value when the nucleon mass is taken
to infinity. This is not true near
threshold\cite{r:GondoloG,r:GriestS}, where $k_3 \rightarrow 0$.
As an example, consider an incident lepton whose energy is the
threshold energy for a nucleon at rest. Then, for those nucleons moving with
$\ctheta < 0$ the effective reaction energy is above threshold. Thus,
after thermal averaging, the threshold should no longer be sharp. Fortunately,
the reaction rates are not dominated by the behavior near threshold, since
phase space vanishes there. The error introduced by the adopted procedure
seems to be acceptably small, as will be discussed later.

Now consider the lepton blocking factor $\fb$.
Assuming that the leptons are in thermal equilibrium (see Dodelson and
Turner\CI{DodelsonT} for a discussion of this point) the blocking factor is
\elab{bf}{
\fb = 1 - n(E_3') = \frac{1}{1+ e^{-E_3'/T_3}},
}
where $E_3' \not= E_3$ is the true energy of the outgoing lepton.
The factor $\fb$ depends only on the energy of the outgoing
lepton. Unfortunately, $E_3'$ is a function of both the scattering
angles and the relative motion of the initial lepton and nucleon, so an
integration over all of phase space is unavoidable.

The relevant corrections to the	blocking factor are derived
in Appendix B and given in \eref{fbs}.
There are four corrections, organized by factors of $\calpha$ and $\ctheta$,
that constitute the blocking factor up to order $1/\mnuc$. The corrections
are normalized by $\fb_0$, the zeroth order term, so that the full
blocking factor is
\elab{fb}{
\fb = \fb_0 (1 + \fbiso + \fbtheta + \fbalpha + \fbalphatheta).
}
$\fbiso$ is first order
in $1/\mnuc$ and should be combined only with the zeroth order part of
$\sigma \vrel$ to produce an effective correction to the cross-section
\elab{sigmafb}{
\gammafb  = \gamma_0 \fbiso,
}
which generates a correction to the rate when integrated over incident
lepton energy. The next term $\fbtheta$ is of order $1/\sqrt{\mnuc}$
and proportional to $\ctheta$. When combined with the $\ctheta$
correction to $\sigma \vrel$ due to thermal averaging,
an effective $1/\mnuc$ correction is produced
\elab{sigmathb}{
\gammathb = \gamma_0 \frac{k_1^2 (E_1(E_3^2 + k_3^2) + E_3 k_3^2)}
   {E_1 E_3 k_3^2 \mnuc (1 + e^{E_3/T_3})} \frac{\Tgamma}{T_3}.
}
Finally there are two
pieces that are proportional to $\calpha$, which must be combined with the
$\calpha$ dependent part of the differential cross-section,
$\sigma_\alpha$. The first piece, $\fbalphatheta$ is proportional
to $\ctheta/\sqrt{\mnuc}$ and must be combined with only the $\ctheta$
part of $\sigma_\alpha \vrel$ to yield a correction
\elab{sigmaalphaa}{
\gammaalphatheta = \gamma_0 \frac{c_a^2 - 1}{1 + 3 c_a^2}
  \frac{2 E_3 k_1^2 + E_1 k_3^2}{3 E_1 E_3 \mnuc (1 + e^{E_3/T_3})}
   \frac{\Tgamma}{T_3}.
}
The second piece, $\fbalpha$ is
angle independent but of order $1/\mnuc$ and is combined only with the
leading piece of $\sigma_\alpha$ to yield a second correction
\elab{sigmaalphab}{
\gammaalpha = \gamma_0 \frac{c_a^2 - 1}{1 + 3 c_a^2}
   \frac{k_1 k_3}{3 E_1 E_3 \mnuc}
 \left( \frac{k_1^2 k_3^2 - (E_3 k_1^2 + 2 E_1 k_3^2) \Tgamma}
   {k_1 k_3 T_3 (1 + e^{E_3/T_3})} -
   \frac{(e^{E_3/T_3} - 1) k_1 k_3 \Tgamma}
       {T_3^2 (1 + e^{E_3/T_3})^2} \right)
}
In the following section, these two terms
are combined to form a single correction, $\gammaalphatot$.

\subsection{Results for the corrected rates}

In the previous two sections six corrections to the weak $\pn$ rates
that are formerly of order $1/\mnuc$ were identified:
$\gammawm$, $\gammarec$, $\gammatha$, $\gammafb$, $\gammathb$,
and $\gammaalphatot$; which should be combined with $\fb_0$
and integrated over $E_1$ to produce corrections to the rates.
\fref{dgamma} shows a plot of $\gamma_i/\gamma_0$ for each of the six
corrections to the reaction $\nun$, at $\Tgamma = 0.8\MeV$.

First, consider the three small terms $\gammathb$, $\gammafb$, and
$\gammaalphatot$, which have all been exagerated by a factor of 100 in the
figure. Clearly, these three terms are much smaller than the other three.
The main reason for this is easy to understand. Due to the $E^2$
dependence of the cross-section and powers of $E$ in phase space,
the rates are dominated by leptons with $E \sim 5 T$. At this point the
blocking factors are small, and corrections to them are even smaller. This
can be seen explicitly in Appendix B, where it is shown that each correction
to $\fb$ carries at least one extra factor of $(1+e^{E_3/T})\M1$.
In addition, the proliferation of terms in the expansions
leading to $\gammawm$, $\gammarec$, and $\gammatha$ is greater than for
the terms associated with the blocking factors. A third
factor suppresses $\gammaalphatot$, namely that it is proportional to
$1 - c_a^2$, which is numerically about a tenth of $1 + 3 c_a^2$ which
comes into the thermal corrections. For all these reasons, the three
small corrections are dropped from most of the discussion that follows.

Now turn to the three larger corrections, beginning with that for weak
magnetism. \fref{wm} shows $\gammawm$ weighted by phase space
considerations to produce a differential interaction rate per baryon,
$d\Gamma_{wm}/dE_1$, that can be found by substituting $\gammawm$ for
$\gamma_0$ in \eref{2to2c}. The scale for this graph should be
compared to \fref{tot0}. The corrections to each reaction are of order
1\% at $\Tgamma = 0.8 \MeV$, but apart from small contributions near thresholds
one can see that there is an almost exact cancellation between the lepton
reactions ($\nun$ and $\ep$) and the anti-lepton interactions
($\eplusn$ and $\nubarp$). This is due to an effective
change in sign for the value
of $c_a$ when considering leptonic and antileptonic scattering, \ie, the
anti-leptonic current is right handed. Thus, although the corrections are
large for each of the individual reactions, the net effect on
nucleosynthesis due to weak magnetism is fairly small. It is not, however,
totally negligible. There are differences in the phase space details
for the different channels, and the neutrino temperature is in fact
less than the electron temperature. As a result,
when the photon temperature is $0.5 < \Tgamma < 2\MeV$
the $\eplusn$ channel is slightly more important than the $\nun$ channel,
and weak magnetism causes a small decrease in
$\Gamma_{\ntop}$. This, in turn, causes a slight increase in $\xn$.
At cooler temperatures, the electron density drops and the $\eplusn$
channel becomes insignificant, but by then the $\nun$ channel is also
small and the weak magnetism corrections are not important then.

Next, consider \fref{rec}, which shows the correction to the differential
reaction rate due to recoil effects, $\gammarec$. Here the sign of the effect
is the same
for all reactions. The final state phase space for the outgoing lepton is
reduced and this causes a reduction in the cross-sections at $T = 0.8 \MeV$
of about 1\%. The magnitude of the reduction increases with temperature.
This can be seen by examining the $\gammarec$ curve in \fref{dgamma}
where the fractional increase in the recoil effect is seen to increase
approximately linearly with energy. When weighted by a thermal
distribution the fractional change in rate will increase with $T$.

Even though all the reactions are affected in a similar way that does not
imply that there will be no effect on nucleosynthesis. Since all the rates are
reduced, freezeout of the neutron-proton ratio will take
place a little earlier, when the neutron abundance is higher. As a result
there will be more helium. Further, the rates are not reduced in
proportion to the zeroth order rates, so there may be a shift in
$\xn/\xp$ even at high temperatures, when the rates are fast.
These effects will be discussed further in section 4.

The third important correction is that due to thermal averaging over the
nucleon momentum, illustrated in \fref{th}.
Here again all reactions are affected in a similar way,
only now the rates are slightly increased. The increase is due to the fact
that the average collision energy is slightly enhanced by the
nucleon motion, and since the cross-sections increase with energy, the
rates increase due to this effect. Comparison of \fref{rec} and \fref{th}
shows that the thermal averaging effect is about 1/3 the effect due to
recoil, so that the net effect of the two processes is
to decrease the reaction rates.

The total reduction in rate arises from integrating over initial lepton
energies. \fref{drate} shows the fractional change in rate,
$\delta = \Delta\Gamma/\Gamma$ for the four scattering reactions
as a function of $\Tgamma$. The curves include all six terms shown in
\fref{dgamma}. The reduction increases nearly linearly with temperature,
although there are deviations at low temperatures. The linear increase is a
consequence of the fact that of the three small parameters, $\me/\mnuc$,
$Q/\mnuc$, and $\sim 5 T/\mnuc$, the latter is by far the largest. The
coefficient, 5, reflects the increase of cross-section and phase space with
initial lepton energy.

Another feature of \fref{drate} is that at high
temperatures the corrections to $\gamnp$ are
either less positive or more negative than the corresponding corrections to
$\gampn$. This is not unexpected since the order $1/\mnuc$
correction to the equilibrium abundance of neutrons should result in a
0.2\% increase in $\xn/\xp$, and this must be reflected by a
change in the rates which maintain equilibrium. At low temperatures two
things happen. First, the neutrino and photon temperatures are no longer
equal, so equilibrium arguments no longer apply. Second, the difficulties
with the threshold behavior in the $\nubarp$ channel become apparent.
Fortunately, the threshold behavior does not become a problem until
$\Tgamma \simlt 0.5\MeV$ and by that time the absolute rate of the
$\nubarp$ reactions is so small (see \fref{inttot0}) that the error to the
correction to $\xn$ is insignificant.

\section{Corrections to Neutron Decay}

As mentioned in the introduction, the $\pn$ rates used in big bang
nucleosynthesis calculations are not usually
calculated from first principles, but
are normalized to the experimental lifetime for neutron decay. Originally
this had the advantage of partially accounting for some of the effects left
out of the calculation, such as the coulomb and radiative corrections. In the
present case, this convention requires us to calculate the recoil
corrections for neutron decay, since those corrections are, in effect,
already included in the numerical BBN codes.

Write the scattering rate for one of the channels as
\elab{gammasc}{
\Gamma_{sc} = \Gamma_{sc,0} (1 + \deltasc),
}
where $\Gamma_{sc,0}$ is the zeroth order scattering rate, and
$\deltasc$ is the $1/\mnuc$ term normalized to $\Gamma_{sc,0}$.
Similarly, the neutron decay rate may be written as
\elab{gammadec}{
\Gamma_{n} = \Gamma_{n,0} (1 + \deltan),
}
where the decay rate is approximated by the sum of
zeroth and first order terms in an expansion in $1/\mnuc$.
The zeroth order scattering and neutron decay rates are related,
schematically, $\Gamma_{sc,0} = B \Gamma_{n,0}$, where $B$ is some
function of temperature and the particle masses. Since the
nucleosynthesis codes are normalized to
the experimental decay rate, but include no recoil corrections they
effectively use a scattering rate $\Gamma'_{sc} = B \Gamma_{n}$. The
correction to the current calculations may then be estimated
\ealab{gammasc1}{
\Gamma_{sc} & = & B \Gamma_{n,0} (1 + \deltasc)  \nonumber \\
     & \approx & \Gamma'_{sc} (1 + \deltasc - \deltan) ,
}
In the last section the various $\deltasc$ were calculated,
implicitly; in this section the corresponding $\deltan$ is evaluated.

For laboratory neutron decay it is only necessary to evaluate the recoil
corrections - there are no thermal averages, nor any blocking factors.
Although weak magnetism affects the angular correlations of the decay
products, its effects drop out of the total decay rate at
first order because the interference term with the axial current is zero
when integrated over leptonic phase space.
This can be used as a check of the calculation.

The decay to three bodies can be put into a form similar to
that for scattering processes,
\elab{decay}{
\Gamma_{\ndec} =
  \int d E_1  \frac{k_2^2}{2 \pi^2} \gammadec ,
}
where $\gammadec$ is identicle in form to that for the $\ntop$
cross-sections but with $s$ evaluated for an `initial' lepton energy
equal to minus the energy of the corresponding lepton in the decay. It is
then straightforward to use $\gammadec = \gamma_0 + \gammarec + \gammawm$.
Graphs of the corresponding differential decay spectra and corrections are
shown in \fref{ndec}. One can see that weak magnetism contributes to the
asymmetry but not to the total decay rate; however, the recoil correction
does reduce the decay rate, by an amount
\elab{dgamman}{
\deltan \approx -0.00201.
}

Noting that we have not included Coulomb
and radiative corrections, the value for the zeroth order neutron halflife
is $\tau_{n,0} = 1/\Gamma_{\ndec} = 964.70 \sec$, while the halflife including
recoil is $\taunrec = 966.66$. In the next section, where the rate equations
are solved for $\xn$, it will be advisable to account for as much of the
Coulomb
and radiative corrections as possible so as to isolate the corrections
due to nucleon mass effects. To do this it should be adequate to adjust both
the neutron decay rate and the scattering cross-sections, by a constant
factor. This can be done easily by increasing the effective Fermi constant
by $(966.66/889.1)^{(1/2)} = 1.0427$.

Wilkinson\CI{Wilkinson} has performed a comprehensive examination of
the corrections to neutron decay. In an effort to obtain a reliable accuracy
at the level of one part in $10^4$, he evaluated all effects that would
plausibly contribute at a level $10\M5$. These include recoil,
weak magnetism, radiative, and coulomb corrections to second order as well
as other small corrections, \eg\ due to the finite size of the nucleons.
Specifically, his Table 4 includes a recoil correction of
$\deltan = + 0.0017$. This result differs
from \eref{dgamman} in magnitude and sign (!), but the difference is due
solely to different definitions of what is meant by the recoil correction.

Wilkinson writes the decay rate as
\elab{wilk}{
\Gamma_n = B' \int_1^{E_0} dE_e
   E_e k_e (E_0 - E_e)^2 (1 + R(E_e, E_0, \mnuc)),
}
where $B'$ is a constant, and $E_0 = Q - (Q^2 - m_e^2)/(2 \mnuc)$ is
the electron endpoint energy, including recoil effects. He then identifies
the recoil correction as
\elab{wilkrec}{
\Delta\Gamma_{n,rec} = B \int_1^{E_0} dE_e
   E_e k_e (E_0 - E_e)^2 R(E_e, E_0, \mnuc),
}
but this does not include the correction to the integral due to the change
in the electron endpoint energy from $Q$ to $E_0$,
\elab{wilkerr}{
\Delta\Gamma'_{n,rec} = B (\int_1^{E_0} dE_e E_e k_e (E_0 - E_e)^2  -
   \int_1^{Q} dE_e E_e k_e (Q - E_e)^2.
}
The change due to the endpoint of integration is small since the integrand
vanishes there in any event, but the decrease in the integrand by
$\approx E_e^2 k_e (Q^2 - m_e^2)/\mnuc$ is significant. Wilkinson includes
this term in his definition of the zeroth order phase space integral,
whereas in the current paper $\Delta\Gamma'_{n,rec}$ is included as part
of the recoil correction. The current nucleosynthesis codes assume that
the change in lepton energy is $E_3 - E_1 = Q$, which is the zeroth order
value for the endpoint conventions used in this paper. Even though
Wilkinson puts $\Delta\Gamma'_{n,rec}$ into the zeroth
order phase space integral, it {\it is} still present in his full phase space
factor, correct to second order in $1/\mnuc$. Therefore, results of neutron
decay based on Wilkinson's work should be valid.

The recoil correction to neutron decay
should not be applied to neutron decay in the early Universe, since the
rate used in the code is the experimentally determined value. There is,
however, a small thermal correction to neutron decay due to the thermal
averaged time dilation factor. The neutron decay rate should be divided
by a factor of $(1 + 1.5 T/\mnuc)$. Since neutron decay is more important
at late times when $T \approx 0.1 \MeV$ this correction, although
technically of order $1/\mnuc$, is numerically quite small.

\section{Estimate of the change in $\yhe$}

All the pieces are now in place to estimate the change in $\yhe$. No
effort will be made in this paper to incorporate the modified $\pn$ rates
in a full nucleosynthesis code. Rather it should be sufficient to examine
the evolution of the neutron fraction down to $T \approx 0.07 \MeV$ with and
without the nucleon mass corrections. The increase in $\yhe$ due to these
corrections is given by twice the increase in the neutron fraction,
$\delyhe = 2 \delxn$.

To perform the evolution, a simplified numerical model of the early Universe
was constructed. One sector included neutrons, protons,
electrons, and photons in thermal equilibrium at a temperature $\Tgamma$.
The other contained three neutrino species
in equilibrium at a temperature $\Tnu$. Account was taken of $e^+ e^-$
annihilation for keeping track of the energy density and the expansion
rate of the Universe, so that in general $\Tgamma \not= \Tnu$. The
effect of different temperatures was included in the rate calculations.

The zeroth order scattering rates, \eref{2to2c}, and the corrections
\eqsref{sigmawm}, \ref{e:sigmarec}, \ref{e:sigmatha},
\ref{e:sigmafb}, \ref{e:sigmathb}, \ref{e:sigmaalphaa}, \ref{e:sigmaalphab}
were calculated on a logarithmic temperature grid and interpolating
functions were created that reproduced
the numerical integration (at new points) to better than a part in $10^4$
over the temperature range, $50\keV < \Tgamma < 10\MeV$. This was done for
each of the four channels. The experimental rate for neutron decay was
modified by the thermal lorentz dilation factor.
The rates for $\invdec$ were inferred using $\Gamma_{\ndec}$ and
the known equilibrium neutron fraction, under the assumption that
$\Tgamma = \Tnu$. Since this channel is numerically unimportant the error
introduced by this procedure is not important.
The reaction rates are then
\ealab{gamcor}{
\Gamma_{sc} &=& \Gamma_{sc, 0} (1 + \deltasc - \deltan) \nonumber \\
\Gamma_{\ndec} &=& \Gamma_{\ndec, 0} (1-1.5 \frac{\Tgamma}{\mn}) \nonumber \\
\Gamma_{\invdec} &=& \Gamma_{\ndec} e^{-Q/T}
}

These rates were used to solve for $\xn$ by
\elab{dxntnu}{
\deriv{\xn}{\Tnu} = - \frac{1}{H \Tnu} \deriv{\xn}{t},
}
with
\elab{dxndt}{
\deriv{\xn}{t} = \Gamma_{\pton} (1-\xn) - \Gamma_{\ntop} \xn,
}
where $\Gamma_{\pton}$ and $\Gamma_{\ntop}$ are sums over the appropriate
reaction rates. $H$ is the expansion rate given by
\elab{exprate}{
H^2 = \frac{8\pi G_N}{3} \sum_i \rho_i(m_i, T_i)
}
where $G_N$ is Newton's constant and $\rho_i$ is the density in species
$i$ calculated for the appropriate mass and temperature. The photon and
neutrino temperatures were derived assuming adiabatic expansion and
totally decoupled neutrinos.

The integration was started at $\Tnu = 10\MeV$. For the zeroth order case the
initial neutron to proton ratio was set to $(\xn/\xp)_0 = e^{-Q/T}$, but for
the calculation with $1/\mnuc$ corrections the initial value was set to
$(\xn/\xp)_1 = e^{-Q/T} (1 + 1.5 Q/\mnuc)$. In fact, the end results
are essentially independent of initial conditions since the reaction rates
are so fast that dynamic equilibrium is quickly achieved.

\fref{xn} shows the resulting $\xn$. The equilibrium values $x_{n,eq}$ are
also shown to illustrate the freezeout of the $\pn$ scatterring reactions,
followed by the slower neutron decay. The breaking of the deuterium
bottleneck is defined, in an {\it ad hoc} way, to occur when $\xn = 0.12$.
This happens at $\Td = 0.071 \MeV$.

The zeroth order and corrected results for $\xn$ are so close that the
difference cannot be shown in \fref{xn}. To bring out the correction,
$x_{n,1} - x_{n,0}$ is plotted as the solid curve in \fref{xn1}.
The maximum correction occurs around freezeout, but is
diminished by neutron decay until the deuteron bottleneck breaks and
the remaining neutrons are cooked into $\Hefour$. The correction to
$\xn$ at this point is $\delxn(\Td) \approx 0.0006$ yielding a correction
to the helium abundance of $\delyhe \approx 0.0012$.

It is interesting that at high temperatures the reaction rates for the model
with nucleon mass corrections do not appear to reproduce the equilibrium
neutron fraction, shown as the dotted curve in \fref{xn1}.
The difference can be
understood as being due to corrections that are second order in $1/\mnuc$
- both in the equilibrium abundance and in the rates.

A test of this
can be done by forming a residual which should vanish through first order
in $1/\mnuc$ when $\Tgamma = \Tnu$,
\elab{comp1}{
\resid = 1 - \frac{\gampn}{\gamnp} \left( e^{-Q/T} (1 + 1.5 Q/\mnuc) \right)
  \sim {\cal O} \left( \frac{1}{\mnuc^2} \right).
}
A graph of $\resid$ is shown as the solid curve in \fref{resid}.
At high temperatures $\resid$ is increasing because the second order
corrections are increasing.
At $T = 10\MeV$ one finds $\resid \approx 10\M4$ which accounts for most
of the difference between $\Delta$ and $\Delta_{eq}$ in \fref{xn1}.
At lower temperatures, $\Tgamma \sim 1\MeV$, there is no problem with the
corrected rates producing corrected equilibrium fractions, rather one only
needs to ascertain that $\resid$ is much less than the individual first
order corrections $\delta_{sc,i}$. Indeed, the residual is much smaller
than the individual corrections (typically a few percent)
for $0.5 \MeV < T < 10\MeV$.

At very low temperatures $\resid$ again becomes significant.
The problem goes back to the poor threshold
behavior of the $\nubarp$ reaction. This was checked by
arbitrarily taking $m_e = 0$, which should alleviate the threshold
problems, and increasing $\mnuc$. In that case,
$\resid$ scaled as $1/\mnuc^2$ across the full temperature range
$0.1\MeV < T < 10\MeV$.

\fref{resid} also shows several other examples of $\resid$ with different
terms included in the rates. The solid curve at the bottom shows $\resid$
in the limit of infinite mass nucleons, and $m_e = 0$. The $10\M7$ level
of the result reflects the accuracy of the numerical integration. The
dotted curves show $\resid$ for $m_e = 0$ in the cases where $\Gamma$
includes a) recoil, b) thermal averaging,
c) recoil and thermal averaging, and d) recoil,
thermal averaging, {it and} the small blocking corrections. For both cases c)
and d), $\resid$ is smaller than in the previous case as more of
the terms necessary to achieve thermal equilibrium are included.
The $10\M4$ magnitude for case d) is indicative of the $1/\mnuc^2$ nature
of $\resid$. Note that it is not necessary to
include the weak magnetism corrections in this analysis, since one can
consistantly imagine another world where $\fwm= 0$, and $\resid$ should
still vanish to second order.

The conclusion of these investigations is that the numerical accuracy of
the approximations and numerical integrations is adequate for
temperatures below $\sim 3\MeV$. The major weakness is the poor threshold
behavior, which induces errors of order the correction in the $\nubarp$
channel for $\Tgamma \simlt 0.5\MeV$. Since this channel is not so
important then, the numerical accuracy of the corrections presented here are
estimated to be about 10\% ($1\sigma$ equivalent). There are also
errors at higher temperatures since the corrections are only first order
in $T/\mnuc$, but these errors are dynamically erased by the fast reaction
rates that persist down to freezeout.

It would be useful to have a simple
approximation for the $1/\mnuc$ corrections, since encoding
the full expression into a nucleosynthesis code and performing the phase
space integrals at each step would be a time consuming exercise.
An approximation, linear in $\Tgamma$ was developed,
\ealab{linear}{
\delta_{\ntop} & = &  -0.00185 - 0.01032 \frac{\Tgamma}{\MeV} \nonumber \\
\delta_{\pton} & = &  +0.00136 - 0.01067 \frac{\Tgamma}{\MeV},
}
which represents averages for the two channels that enter into
the forward or back reactions. As such these may be readily applied to the
polynomial formulae used in Wagoner's code to approximate the $\ntop$
and $\pton$ reaction rates. Before doing this one must separate out those
pieces due to neutron decay and inverse decay and treat them on a separate
footing, as in \eref{gamcor}. The approximations in \eref{linear} do not
include the correction to the neutron lifetime, so this must be added in
separately.

The result of carrying out this procedure for the simplified model of the
early Universe used in this paper is shown as the dashed curve in \fref{xn1}.
The solution for $\xn$ matches that derived from direct integration
of the rates to better than 10\% for temperatures less than $2 \MeV$.
This is comparable to the estimated uncertainty in the calculation of
the rates due to the improper treatment of the threshold effects.
The parameters in \eref{linear} were
chosen by fitting the $\pton$ reactions in the temperature range
$0.7 < \Tgamma < 2 \MeV$, and the $\ntop$ reactions in the range
$0.3 < \Tgamma < 2 \MeV$. These ranges cover freezeout for the different
channels and avoid, for the most part, sensitivity to the threshold
behavior of the rates.

Finally, to isolate the effects of weak magnetism, $\xn$ was calculated
with a set of rates where $\fwm$ was set to zero. The resulting increase in
$\xn$ was $0.00045$ instead of $0.0006$; \ie about 1/4th of the net increase
in $\yhe$ can be attributed to weak magnetism. The bulk of this
contribution comes at $0.5 < \Tgamma < 2 \MeV$ where the $\eplusn$ channel
is slightly more important than the $\nun$ channel because of kinematics
and also because $\Tgamma$ is slightly greater than $\Tnu$.

It doesn't really make sense to perform a similar calculation to try and
isolate the recoil {\it vs.} the thermal averaging corrections. The point of
the
analysis of the residual $\resid$ is that both are necessary to achieve a
sensible thermodynamic result if one were to take $\Tgamma= \Tnu$. Even
so, including just recoil corrections to the rates, leads to a change in
$\xn$ of just 0.0002. This is somewhat surprising since the recoil
corrections were larger than and of the opposite sign to the thermal
averaging corrections. Based on this, one might have expected the recoil
corrections to give a correction to $\xn$ of order $\sim 0.0008$, which
would be partly compensated by the thermal averaging corrections. This is
not the case. An explanation can be found in the details of \fref{tot0}
and \fref{rec}, where the corrections can be seen to be not simply
proportional to the zeroth order rates.

\section{Discussion}

The main point of this paper is that the primordial helium abundance
predicted by big bang nucleosynthesis calculations should be increased by
$\delyhe \approx 0.0012$. It is difficult to attach a firm level of uncertainty
to this number, but the results displayed in \fref{resid} and the accompanying
text suggest that an uncertainty of 10\% should be inferred.

This is a significant correction, but does not
dramatically alter the conclusions that may be drawn from studies of BBN.
Consider the changes implied for the baryon density of the Universe as
inferred from nucleosythesis calculations. Walker and
Kernan\cite{r:Kernan,r:Walker} have recently analyzed the uncertainties
in the big bang helium calculation, but they do not include the
corrections due to nucleon mass effects. Adapting their result to include
the results presented here, the primordial helium abundance for
the standard cosmology with three neutrino species is
\elab{helium}{
\yhe = 0.2410 + 0.0117 \ln(\frac{\eta_{10}}{3}) \pm 0.0017 \pm 0.0002,
}
where $\eta_{10} = 10^{10} n_B/n_\gamma$ parameterizes the baryon density.
Without the $1/\mnuc$ corrections, the first coefficient would be 0.2398,
instead of 0.2410. In \eref{helium}, the
first uncertainty represents a $2\sigma$ error due to uncertainties in
the nuclear reaction network, of which ``80-90\%'' is due to uncertainty in
measurements of the neutron decay rate. The second uncertainty allows
for some of the smaller corrections to the weak interaction rates,
for example, the deviation of the neutrino spectrum from thermal
equilibrium. The nucleon mass corrections,
approximately $0.0012$ in $\yhe$, are equivalent to about a $1.5\sigma$ shift
in the neutron decay rate, and are much larger than any other known
uncertainty in calculating the weak rates.

It is difficult to determine the primordial abundance of $\Hefour$ through
direct observation due to a) the inert nature of neutral helium, b)
chemical pollution through stellar burning, and c) the high accuracy of
the measurement required - better than 1\% is desired. Walker, \etal\
suggest a primordial abundance in the range $0.22 < \yhe < 0.24$. The
limits are suggestive of 95\% confidence levels, but there is no
statistically rigorous upper bound to the helium abundance. For the sake
of argument then, take 0.24 as the upper limit and allow the
uncertainty due to the neutron lifetime in \eref{helium}
to be favorable at the $2\sigma$ level, \ie, allow the two
uncertainties to add $-0.0017$ to the helium
abundance\footnote{
In fact the $\pm 0.0017$ was derived using a $3.5 \sec$
$1\sigma$ uncertainty in the neutron lifetime\CI{Kernanpc},
where the current particle
data book uncertainty is $2.1 \sec$, so the uncertainty due to the reaction
network should become $\sim \pm 0.0014$. The corresponding decrease in the
maximum allowed value of $\eta_{10}$ would decrease from 3.18 to 3.10.}.
These constraints require $\eta_{10} < 3.18$. Without the 0.0012
correction due to nucleon masses the corresponding number is
$\eta_{10} < 3.53$. These numbers should be compared with
the constraints derived from comparisons of observations of
D, $\Hethree$, and $\Liseven$ and BBN
calculations, $2.8 < \eta_{10} < 4.0$.

Taken at face value, a substantial portion (but not all) of the allowed
parameter space for $\eta_{10}$ is eliminated by the nucleon mass
correction; however, one should always keep in mind the difficulties of
helium observations. If the upper limit were $\yhe < 0.245$ there would be
no significant constraint from the consideration of $\Hefour$. On the
other hand, the discussion in the previous paragraph was based on two
separate favorable assumptions, both at the $2\sigma$ level: a) allowing
$\yhe = 0.24$, and b) taking the neutron lifetime to be near the lower end of
the allowed range. Dropping either of these assumptions from the favorable
to the neutral category eliminates any allowed values of $\eta_{10}$.

Another use of the primordial helium abundance is to constrain the energy
density at the time of nucleosynthesis. This is often parameterized by the
number of neutrino species, $\delyhe = 0.012 (N_\nu-3)$. The $1/\mnuc$
corrections are equivalent to $0.1$ neutrino species. Again, belief in
constraints placed on particle physics models depends upon one's faith
in the helium observations.

Over the years, there have been several papers written which treat recoil
corrections in $\ntop$ processes. In addition to the Wilkinson paper on
neutron decay, Fayans\CI{Fayans} and Vogel\CI{Vogel} have studied recoil
and weak magnetism corrections to the $\nubarp$ reaction in the context of
laboratory neutrino oscillation experiments. The results here
are in agreement with Fayans for both recoil and weak magnetism. There is
also agreement with Vogel concerning weak magnetism. It is more difficult
to compare to Vogel's results for the recoil correction, since he gives
the correction in terms of the final state lepton energy, whereas the
results in this paper express the corrections in terms of the initial
lepton energy. I know of no paper which deals with the thermal
corrections or the corrections to the blocking factors that are relevant
for the big bang nucleosynthesis scenario.

\noindent
\underline{Acknowledgements}
While this work was underway I became aware of the work by
Kernan\CI{Kernan} and Walker\CI{Walker}, who were also beginning
to look into the question
of recoil corrections. I am indebted to them for sharing the details of
their previous work and thoughts about the issues presented here.
I would also like to thank S.M.~Barr, P.~Vogel,
J.~Engel and E.W.~Kolb for useful discussions.
This work was partially supported by DOE grant DE-AC02-78ER05007,
and by the University of Delaware Research Foundation.

%
% Hack for equation numbering of appendicies
%
\def\theequation{\ksection.\arabic{equation}}
\section*{Appendix A: The squared matrix element}
\def\ksection{A}              % kloodge for equations
\setcounter{equation}{0}
\def\fsig{\fwm}
\def\fa{g_A}
\def\fv{g_V}
For completeness, here is the spin summed squared matrix element for the
Lagrangian in \eref{lweak} and \eref{hadcur}, to leading order in $\Gfermi$.
The terms are grouped by coupling constant and expressed in terms of
relativistic invariants. The invariant $u$ has been eliminated in favor of
$s$ and $t$ and particle masses. The particle identifications are;
1: incoming lepton, 2: incoming baryon, 3: outgoing lepton, 4: outgoing
baryon. In this expression the vector and axial couplings are given
explicitly as $\fa$ and $\fv$, instead of specifying the ratio
$c_a = \fa/\fv$, as in the text.
\ealab{msquare}{
{\cal M}^2 & = &  \Gfermi^2 ( 4 \fps^2 (t-(m_2 - m_4)^2)
   ((m_1^2 + m_3^2) t - (m_1^2 - m_3^2)^2) \nonumber \\
&+ &
    8 \fa^2 (2 m_1^2 m_3^2 + 2 m_2^2 m_4^2 +
           2 s^2 + ((m_2 + m_4)^2 - t)(m_1^2 + m_3^2 - t) \nonumber \\
& & - 2 s (m_1^2 + m_2^2 + m_3^2 + m_4^2 - t)) \nonumber \\
&+ &
 8 \fv^2 (2 m_1^2 m_3^2 + 2 m_2^2 m_4^2 + 2 s^2
   + ((m_2-m_4)^2 - t) (m_1^2 + m_3^2 - t)
  \nonumber \\
& & - 2 s (m_1^2 + m_2^2 + m_3^2 + m_4^2 - t)) \nonumber \\
&+ &
8 \frac{\fsig \fv}{\mnuc} (
 (m_2 - m_4) (m_2^2 (m_1^2 - t) -  m_4^2 (m_3^2 - t) - (m_1^2 - m_3^2) s)
\nonumber \\
& &   + m_4 (-m_1^2 + t) (m_1^2 - m_3^2 + t) +
       m_2 (-m_3^2 + t) (-m_1^2 + m_3^2  + t) ) \nonumber \\
&+ &
 16 \fa \fv ( (m_1^2 - m_3^2)(m_4^2 - m_2^2)
 + (m_1^2  + m_2^2  + m_3^2  + m_4^2  - 2 s  - t) t) \nonumber \\
&+ &
 8 \frac{\fa \fsig}{\mnuc} ( (m_1^2 - m_3^2)(m_4^2 - m_2^2)
 + (m_1^2  + m_2^2  + m_3^2  + m_4^2  - 2 s  - t) t) (m_2+m_4) \nonumber \\
&+ &
 16 \fa \fps ( (m_2 - m_4)(m_2^2 m_3^2 - m_1^2 m_4^2)
  + (-m_1^2 + m_3^2) (-m_2 + m_4) s \nonumber \\
  & & + m_1^2 m_4 (m_1^2 - m_3^2 - t) - m_2 m_3^2(m_1^2 - m_3^2 + t ))
\nonumber \\
&+ &
 \frac{\fsig^2}{\mnuc^2} (2 m_2^4 (m_1^2 - m_3^2 - t) + 2 m_4^4 (-m_1^2 + m_3^2
- t) -
      (m_1^2 + m_3^2)^2 t + (m_1^2 + m_3^2) t^2 \nonumber \\
& & + m_4^2 ( (m_1^2 + m_3^2)^2 - 4 m_1^4 - 3 m_3^2 t + m_1^2 t + 2 t^2)
\nonumber \\
& & + m_2^2 ( (m_1^2 + m_3^2)^2 - 4 m_3^4 - 3 m_1^2 t + m_3^2 t + 2 t^2)
\nonumber \\
& & - 2 m_2 m_4 ( (m_1^2 - m_3^2)^2 + (m_1^2 + m_3^2 - 2 t)t) \nonumber \\
& & + 4 s ( (m_1^2 - m_3^2)(m_4^2 - m_2^2)
  + (m_1^2 + m_2^2 + m_3^2 + m_4^2 - t) t) - 4 s^2 t) ).
  }
When evaluating the weak magnetism coupling numerically, $\mnuc$ was set
equal to the initial hadron mass, $m_2$.

\section*{Appendix B: Corrections due to the final state occupation number}
\def\ksection{B}              % kloodge for equations
\setcounter{equation}{0}

In this appendix, the corrections to the final state blocking factor,
$\fb$ are derived. The corrections can be put into four categories based on
their angular dependences, which also determines how they will be combined
with various corrections to $\sigma \vrel$. There are corrections that are
independent of both $\calpha$ and $\ctheta$, corrections that are linear
in $\calpha$ or in $\ctheta$, and terms that are
proportional to $\calpha \ctheta$.
Let us refer to these four terms as $\fbiso$, $\fbalpha$, $\fbtheta$, and
$\fbalphatheta$. There are also terms that are
proportional to $\cos^2 \theta$, but
these are already of order $1/\mnuc$ and so may be averaged over $\theta$
immediately, \ie\ $\cos^2\theta \rightarrow 1/3$, and included in $\fbiso$
or $\fbalpha$.

Denote the true value of the outgoing lepton energy
by $E_3'$, its value in the infinite nucleon mass case by
$E_3$, and the difference by $\epsilon$; \ie\ $E_3' = E_3 + \epsilon$.
The blocking factor can then be written as
\ealab{block1}{
\fb & = & \frac{1}{1+e^{E_3/T_3} e^{\epsilon/T_3}} \nonumber \\
  & = & \frac{1}{1+a}
   \left(1 + \frac{a}{1+a}\frac{\epsilon}{T_3}
     + \frac{a (a-1)}{2 (1+a)^2}\frac{\epsilon^2}{T_3^2} \cdots \right)
\nonumber \\
     & \approx & \fb_0(1 + \fbtheta + \fbalphatheta + \fbiso + \fbalpha)
}
where $a = e^{E_3/T_3}$, and the last equation defines the normalization
to the corrections. In \eref{block1} the blocking factor has been
expanded to second order in $\epsilon$ in recognition of the fact that
the energy correction will have terms of order
$\vnuc \sim (T/\mnuc)^{1/2}$ which need to be included to second order.

The correction to the outgoing lepton energy, $\epsilon$,
can be derived by a series of Lorentz transformations. Start
by choosing a coordinate system where the nucleon moves along the x-axis,
then a) boost by $\beta$ to the rest frame of the nucleon,
b) rotate by $\theta_1$ so that the lepton
lies along the x-axis, and c) boost by $\betacm$ to the center of mass frame.
After scattering, the inverse Lorentz transformation is
applied and the final state lepton energy is determined as a function of the
scattering angles and initial parameters. The result of this procedure is
\ealab{e3}{
E_3' & = & (1 - \beta \betacm \ctheta_1) \gamma \gammacm \Ecm \nonumber \\
 & & + \calpha (-\betacm + \beta \ctheta_1) \gamma \gammacm \kcm -
  \beta \gamma \cos\psi \sin\alpha \sin\theta_1 \kcm
}
where $\Ecm = (s + m_3^2 - m_4^2)/(2\sqrt{s})$ is the final lepton energy
in the center of mass frame for the collision and $s$ is the usual
relativistic invariant. Using
$s = m_1^2 + m_2^2 + 2 E_{1,cm} m_2$, where
$E_{1,cm} = E_1 \gamma (1 - \beta v_1 \ctheta)$ is the initial lepton
energy in the center of mass frame, produces a result
in terms of the initial energies and momenta in the fluid rest
frame. The other quantities used here are, $\kcm$ - the three momentum
corresponding to $\Ecm$, $\psi$ - the azimuthal scattering angle,
and $\gamma$ and $\gammacm$ - the relativistic $\gamma$ factors
corresponding to the two boosts. To make contact with notation in the
rest of the paper, $\vnuc \equiv \beta$.

To leading order in $\mnuc$, $\betacm = -k_1/m_2$ and $\gammacm = 1$.
Further, $\beta \sim 1/\sqrt{\mnuc}$,
so all terms of order $\beta^3$, or $\beta \betacm$ may be dropped.
This allows \eref{e3} to be reduced to
\elab{e31}{
E_3' =  \gamma \Ecm +   \calpha (-\betacm + \beta \ctheta_1) \kcm -
  \beta \cos\psi \sin\alpha \sin\theta_1 \kcm.
}
Next, extract the leading piece from $\gamma \Ecm$,
\ealab{ecm1}{
\gamma \Ecm & = & \gamma (E_{1,cm} + Q - \frac{k_1^2 + k_3^2}{m_2}) \nonumber
\\
   & = & E_3 + E_1 (\beta^2 - \beta v_1 \ctheta) -
       \frac{k_1^2 + k_3^2}{m_2},
}
where after extracting a $1/\mnuc$ term the initial lepton momenta, $k_1$
and $k_3$, may be used without further correction. Eliminating $E_3$
from \eref{e31} and \eref{ecm1} gives,
\ealab{eps1}{
\epsilon & = & (E_1 (\beta^2 - \beta v_1 \ctheta) -
    \frac{k_1^2 + k_3^2}{m_2}) \nonumber \\
& & + \calpha (-\betacm + \beta \ctheta_1) \kcm -
    \beta \cos\psi \sin\alpha \sin\theta_1 \kcm \nonumber \\
& \equiv & \epsilon_1 + \epsilon_2 + \epsilon_3,
}
where the $\epsilon_i$ are defined respectively by the previous line of
\eref{eps1}.

To determine all relevant contributions to the $\fb$ one must include both
first and second order terms in $\epsilon$. The $\epsilon_3$ term
contributes only to $\fbiso$ after being squared and angle averaged. Since
such a term is second order in $\beta$, one may use $\theta_1 = \theta$
and $\kcm = k_3$. The other two terms are more complicated. There are
contributions to both $\fbalpha$ and $\fbalphatheta$ from $\epsilon_2$ and
from the product $\epsilon_1 \epsilon_2$. In the linear contribution from
$\epsilon_2$ one must keep $\kcm$ and $\ctheta_1$ to sufficient accuracy;
$\kcm \approx k_3 - (\beta \ctheta E_3 k_1)/k_3$ and
$\ctheta_1 \approx \ctheta - \beta(1-\cos^2\theta) E_1/ k_1$. For all the
other terms it is sufficient to take $\kcm = k_3$ and $\theta_1 = \theta$.
The $\epsilon_1^2$ and $\epsilon_2^2$ terms contribute to $\fbiso$,
whereas the linear term in $\epsilon_1$ contributes to both $\fbiso$ and to
$\fbtheta$.

The zeroth order blocking factor and four corrections are then,

\ealab{fbs}{
\fb_0 & = & \frac{1}{1 + e^{E_3/T_3}} \nonumber \\
\fbiso & = & \frac{-k_1^2 - k_3^2 + 3 (E_1 + E_3) \Tgamma}
   {2 \mnuc T_3 (1 + e^{E_3/T_3})} +
  \frac{(k_1^2 + k_3^2) (1 - e^{E_3/T_3}) \Tgamma}
     {2 \mnuc T_3^2 (1 + e^{E_3/T_3})^2}
      \nonumber \\
\fbtheta & = & \frac{(3 \Tgamma)^{1/2} \ctheta k_1}
   {\mnuc^{1/2} T_3 (1 + e^{E_3/T_3})} \nonumber \\
\fbalpha & = & \calpha \left(
  \frac{k_1^2 k_3^2 - (E_3 k_1^2 + 2 E_1 k_3^2) \Tgamma}
    {k_1 k_3 \mnuc T_3 (1 + e^{E_3/T_3})} -
  \frac{(e^{E_3/T_3} - 1) k_1 k_3 \Tgamma}
       {\mnuc T_3^2 (1 + e^{E_3/T_3})^2}
\right) \nonumber \\
\fbalphatheta & = & \frac{(3 \Tgamma)^{1/2} \calpha \ctheta k_3}
      {\mnuc^{1/2} T_3 (1 + e^{E_3/T_3})}.
}

%\BI{hadronicBBN} For a discussion of entropy considerations see,
%\APJ{R.J.~Scherrer and M.S.~Turner}{331}{19}{88};
%but if hadronic
%decay channels are available, the constraining temperature is more
%like $6\MeV$. See,
%\PRD{ M.H. Reno and D. Seckel} {37} {3441} {88};
%\NPB{G.~Lazerides, R.~Schaefer, D.~Seckel, and Q.~Shafi}{346}{193}{90}.
%\BI{Saleem} \PRD{S.~Saleem}{36}{2602}{87}.
%\BI{ScherrerCRS} For example,
% \PRD{R.J.~Scherrer, J.~Cline, S.~Raby, and D.~Seckel}{44}{3760}{91}.

\newpage
\fig{tot0}{The zeroth order differential rates $\deriv{\Gamma_0}{E_1}$
for the $\pn$ reactions, see \eref{2to2c}. Results are shown for
$\Tgamma = 0.8 \MeV$, and $\Tnu = 0.7926$.}{.1in}

\fig{inttot0}{The zeroth order scattering rates for $\pn$.
The neutron decay rate and the expansion rate, $H$, are shown as bold
solid lines.}{.1in}

\fig{dgamma}{The six corrections $\gamma_i$ normalized to
$\gamma_0$ for the reaction $\nun$, as a function of initial
neutrino energy.}{.1in}

\fig{wm}{The weak magnetism correction for the four scattering reactions.
Since the $c_a \fwm$ interference changes sign for reactions with
antileptons rather than leptons, the sum of the corrections is near zero.
$\Tgamma = 0.8 \MeV$.}{.1in}

\fig{rec}{The same as \fref{wm}, but for the recoil corrections,
$\gammarec$. Here the corrections are all negative, since the phase space
for the outgoing lepton is reduced due to the recoil of the nucleon.}{.1in}

\fig{th}{The same as \fref{wm}, but for $\gammath$, the correction due to
thermal averaging over the initial nucleon distribution.}{.1in}

\fig{drate}{The sum of the $1/\mnuc$ corrections to the $\pn$ scattering
rates, expressed as a percentage of the zeroth order rates.}{.1in}

\fig{ndec}{The neutron decay spectrum and the $1/\mnuc$ corrections.}
{.1in}

\fig{xn}{The fraction of baryons in neutrons, $\xn = n_n/n_B$,
as a function of the photon temperature, $\Tgamma$. The deuterium
bottleneck is defined by where $\xn = 0.12$. The equilibrium abundance is
shown as a dotted line.}
{.1in}

\fig{xn1}{The change in $\xn$ due to the inclusion of nucleon mass
corrections. The solid curve shows the result using the full formulae for
the corrections. The dashed curve, $\Delta_{lin}$, shows the result using
the linear approximation to the correction from \eref{linear}.
The correction to the equilibrium abundance is shown as a dotted line.}
{.1in}

\fig{resid}{The residual, $\resid$ (see text), as a function of temperature.
The `dips' occur when $\resid$ goes through zero as two
terms cancel.}
{.1in}

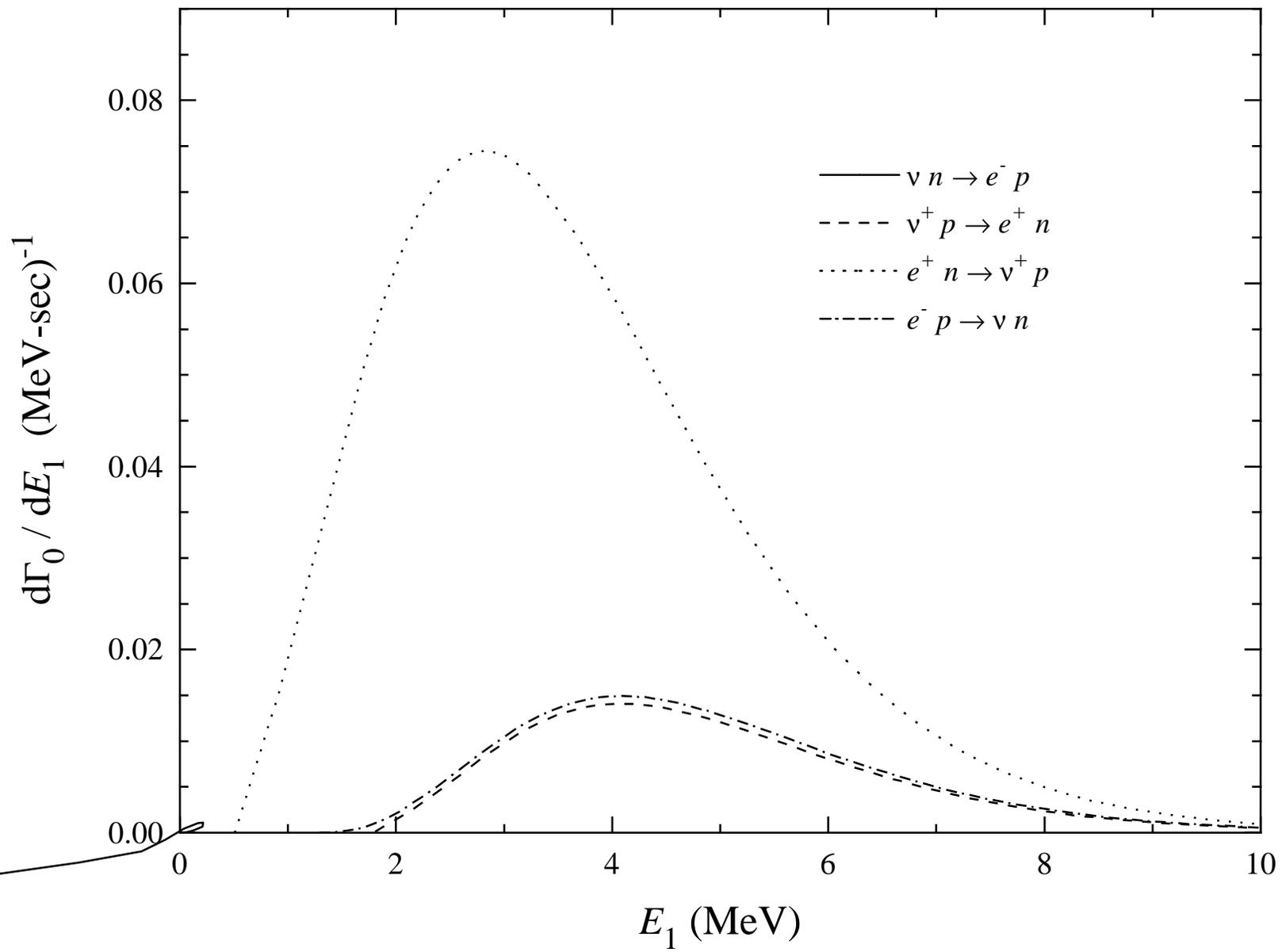
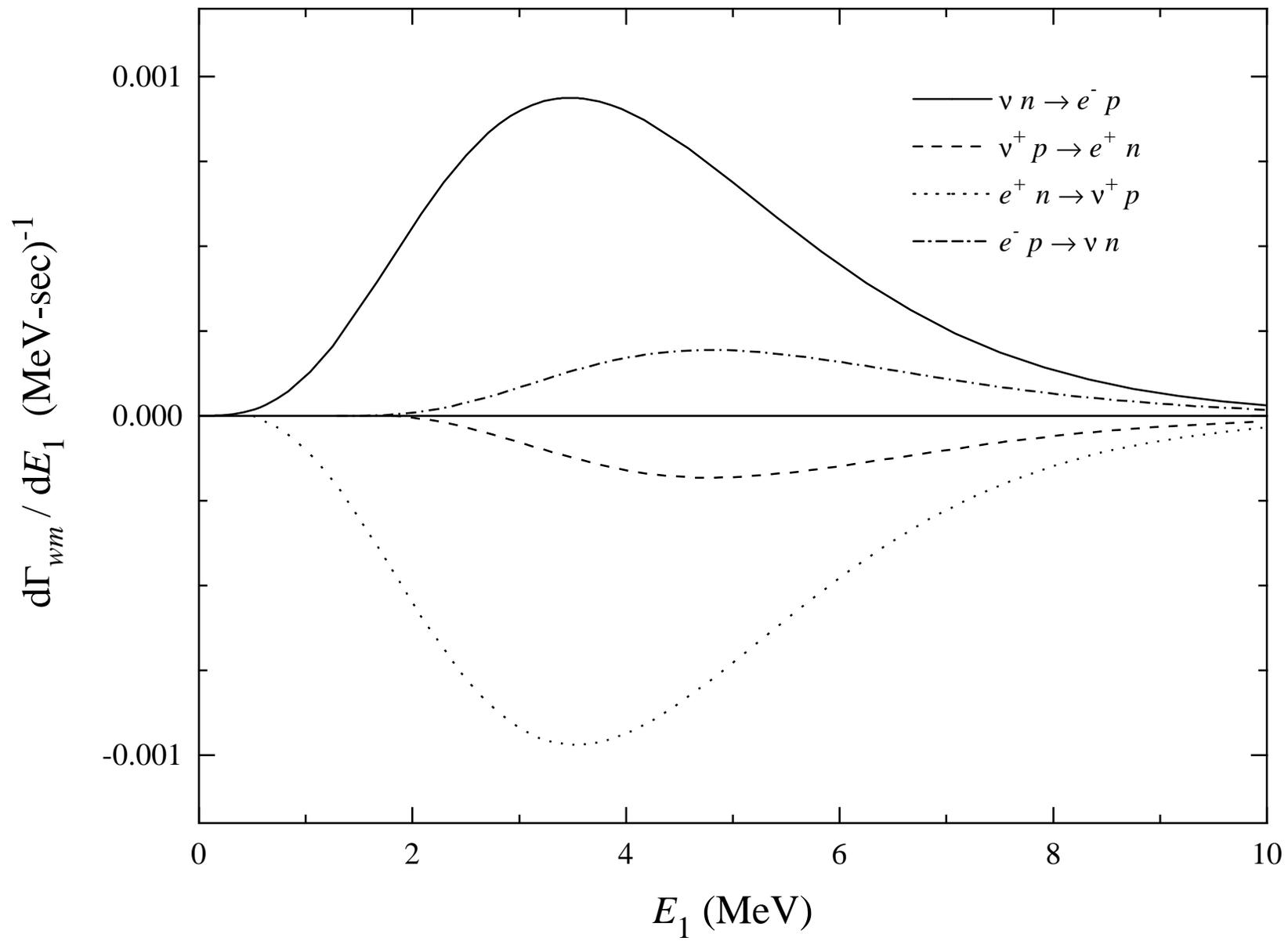
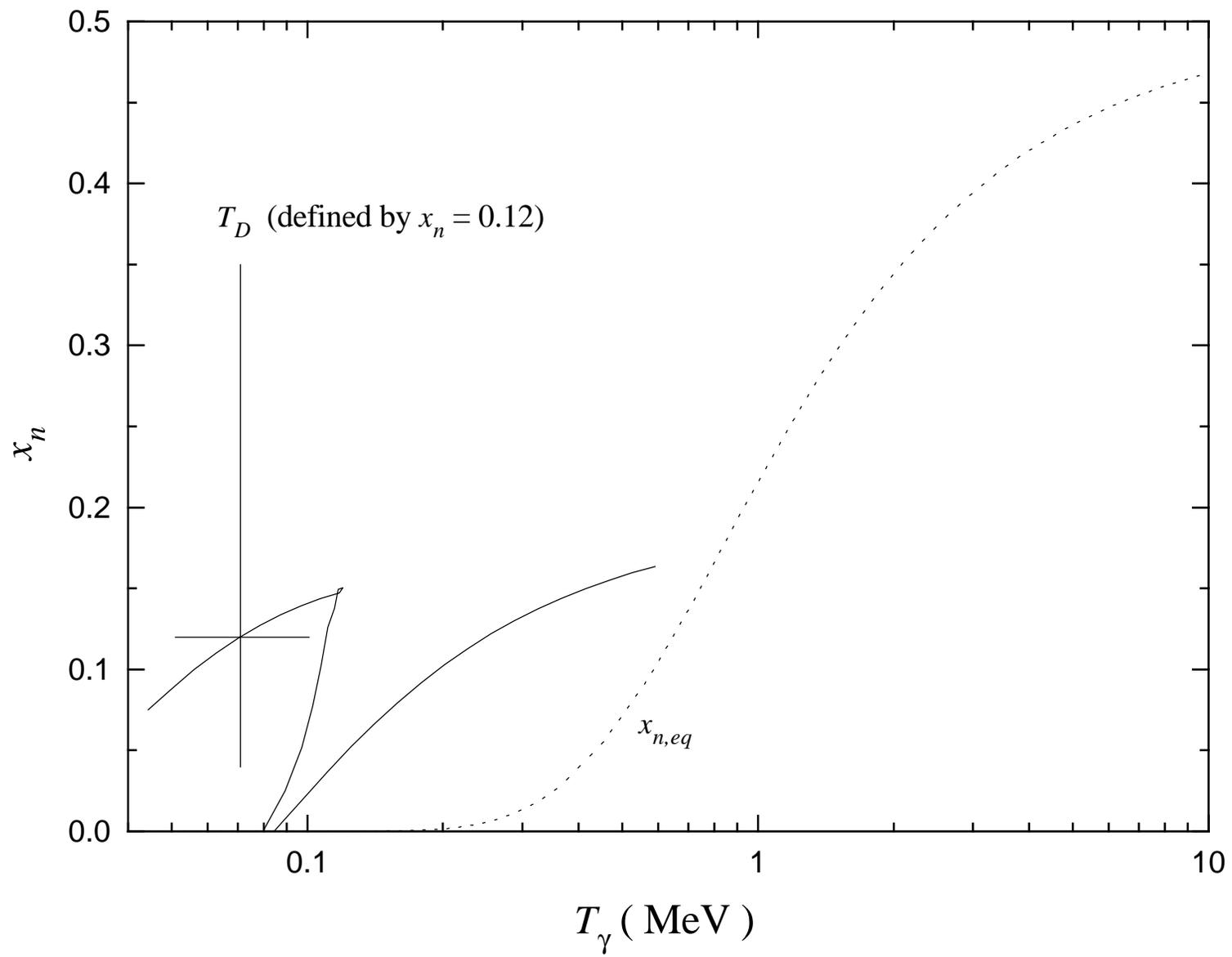
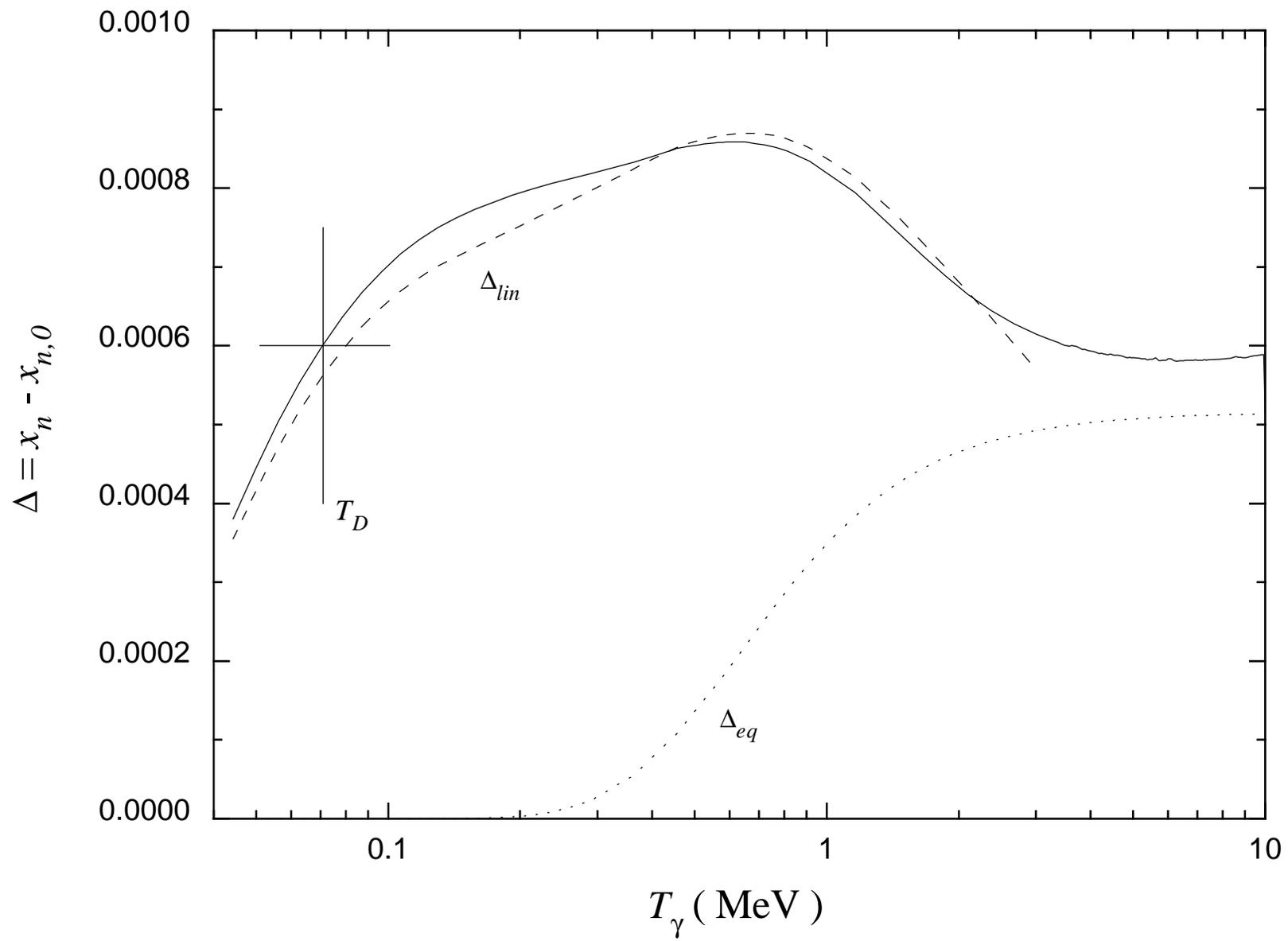
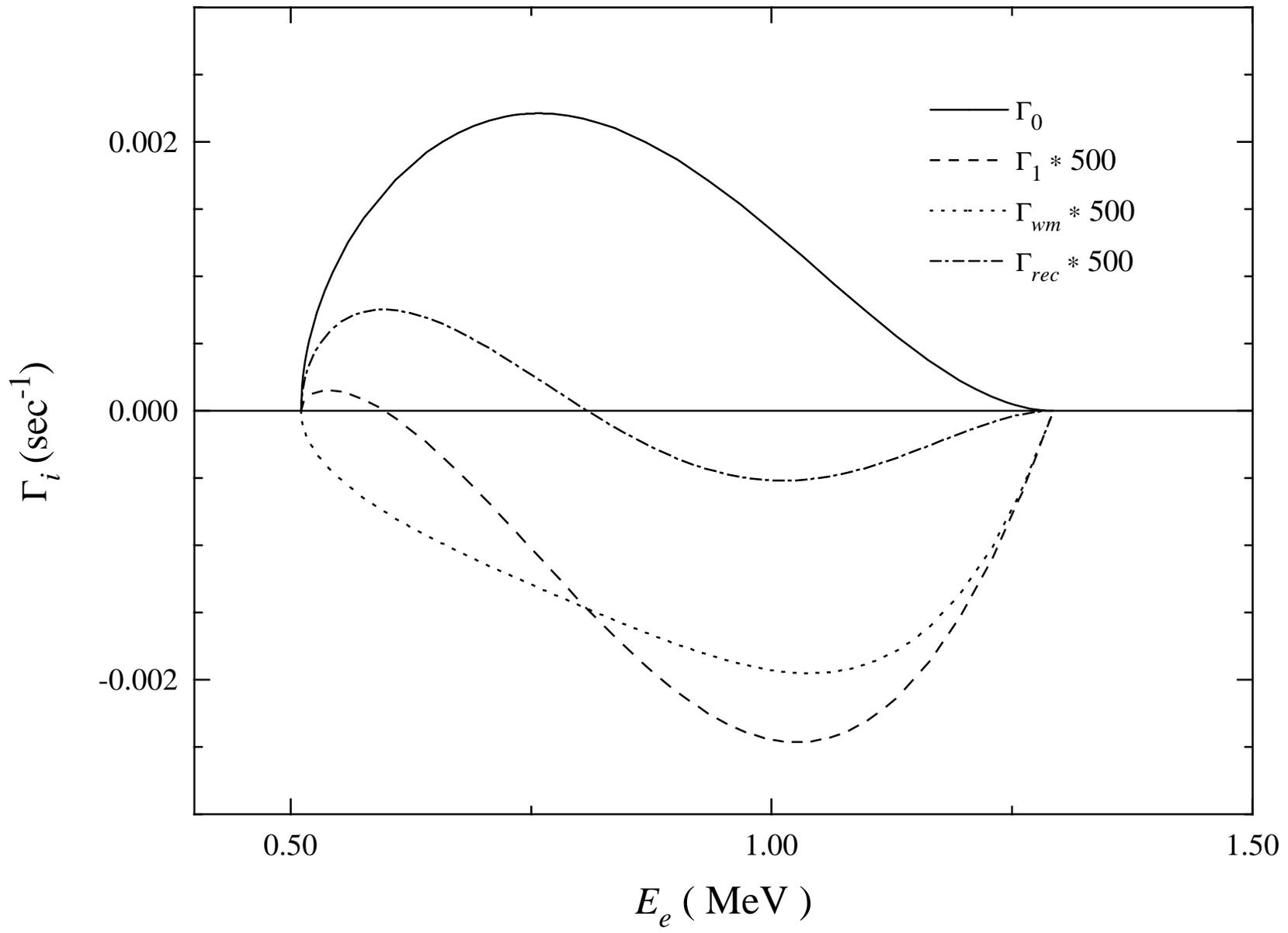
\end{document}